\documentstyle[times,epsfig]{jaa}

\begin{document}

\title[Lessons for Asteroseismology]
{Lessons for Asteroseismology from White Dwarf Stars} 

\author[T.~S. Metcalfe]
 {Travis S. Metcalfe\thanks{e-mail: travis@hao.ucar.edu}\\
 High Altitude Observatory, NCAR, P.O. Box 3000, Boulder CO 80307 USA}

\maketitle

\begin{abstract}
The interpretation of pulsation data for Sun-like stars is currently
facing challenges quite similar to those faced by white dwarf modelers ten
years ago. The observational requirements for uninterrupted long-term
monitoring are beginning to be satisfied by successful multi-site
campaigns and dedicated satellite missions. But exploration of the most
important physical parameters in theoretical models has been fairly
limited, making it difficult to establish a detailed best-fit model for a
particular set of oscillation frequencies. I review the past development
and the current state of white dwarf asteroseismology, with an emphasis on
what this can tell us about the road to success for asteroseismology of
other types of stars.
\end{abstract}


\section{Introduction}

For the past six years, I have been trying to improve the way that we
match theoretical models to observations of pulsating white dwarfs. I have
made some progress transforming the art of model-fitting from a hands-on
procedure to something more objective, more global, and more automated.
The impact of this approach on the analysis of pulsating white dwarfs over
the past few years suggests that it might also be able to improve our
seismological modeling of other types of stars.

I am now interested in applying this same approach to the relatively
recent (and much more difficult) observations of solar-like oscillations
in other stars. Reading through the recent literature on this subject, I
have been impressed by how the current state of model-fitting for these
stars resembles the state of white dwarf modeling ten years ago. With this
in mind, my goal here is to share some of the lessons I've learned while
working on white dwarfs, with the hope that they will be useful for the
future analysis of Sun-like stars.

Before I go any further, I'd like to make it clear that I do not think the
current analysis methods for Sun-like stars are {\it wrong}. I do think
that the current methods are limited in various ways, and these
limitations may prevent us from learning all that we can learn from the
observations. I hope to demonstrate this possibility with some anecdotes 
about our recent progress modeling the pulsating white dwarfs.

\section{Similarities \& Differences}

In case you haven't noticed, the Hertzsprung-Russel diagram is absolutely
filled with pulsating stars \nocite{jcd98} (see Christensen-Dalsgaard
1998, Fig.~1). We believe that solar-like oscillations should be excited
in just about any star with a convection zone. The basic idea is that
convection generates sound waves at all frequencies (``white noise'', like
static on a radio). Some of these frequencies are resonant inside the
spherical cavity of the star and become long-lived global pulsations
(standing waves). Such pulsations have recently been detected in several
main-sequence, subgiant, and giant stars. The oscillation amplitudes and
the frequency of maximum power in these stars agree reasonably well with
our theoretical expectations \nocite{bk03} (see Bedding \& Kjeldsen 2003
for a recent review).

There are several classes of pulsating white dwarfs, almost equally
spaced in $\log\ T_{\rm eff}$. In each case, the pulsations seem to be
excited by a similar mechanism (extra opacity from a near-surface partial
ionization zone) but the surface chemical composition can be hydrogen,
helium, or something else. Compared to other types of stars, white dwarfs
are relatively simple from a physical perspective. Nuclear fusion in no
longer active, and potentially complicating factors like fast rotation or
strong magnetic fields are not a problem for most of these stars
\nocite{win98} (Winget 1998).

The spherical symmetry of gravity allows us to describe stellar pulsations
with spherical harmonic functions, the same mathematics that we use to
describe electron orbitals in quantum mechanics. We can assign each
pulsation mode three quantum numbers: an $\ell$ and $m$ value to describe
the pattern on the visible surface, and an $n$ value (sometimes called
$k$) for different radial overtones. For observations of stars, only modes
with relatively low values of $\ell$ can be detected, because the many
bright and dark areas tend to cancel each other out for higher-$\ell$
modes.

Asteroseismology of white dwarfs has advanced more quickly than for other
types of stars, in part because the pulsations are much less subtle. The
total observed light variation is typically 10\% or more, on very
convenient timescales of 5-15 minutes. The light variations caused by
solar-like oscillations are considerably smaller, so scintillation in the
Earth's atmosphere severely limits our ability to detect them. However,
the pulsations also cause variations in the spectral line profiles and
equivalent widths, allowing us to detect them from the ground with
time-series spectroscopic measurements.

\setcounter{figure}{0}
\begin{figure}
\begin{center}
\epsfig{file=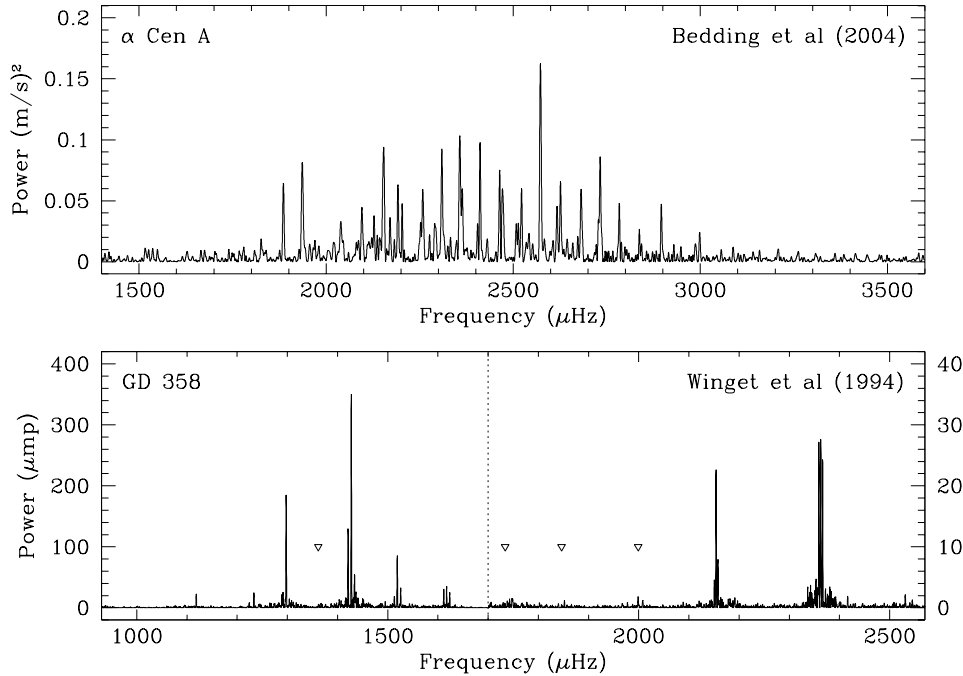,width=9.0cm,angle=270}
\caption{Power spectra of time-series measurements to detect solar-like 
oscillations in $\alpha$~Cen~A published recently (top), compared to
pulsations observed in the white dwarf GD~358 published ten years earlier 
(bottom). The dotted line in the bottom panel divides the regions where 
the left and right vertical scales are applicable, and the open triangles
mark the frequencies for several of the lower amplitude modes.\label{fig1}}
\end{center}
\end{figure}
\nocite{win94,bed04}

From an observational standpoint, Figure~\ref{fig1} illustrates just how 
similar the current status of Sun-like stars is to the situation for white 
dwarfs ten years ago. In both panels we see a well defined range of 
excited frequencies. For Sun-like stars we see a series of almost equally 
spaced frequencies, which is characteristic of pressure (p-) modes. For 
white dwarfs the peaks are unevenly spaced in frequency, but evenly spaced 
in period. This is characteristic of the gravity (g-) modes.

From a theoretical standpoint, the observed pulsation periods are
determined by the values of the sound speed (or Lamb frequency, $L$) and
the buoyancy frequency ($N$) from the center of the star to the surface.  
The p-modes are excited at frequencies higher than both $L$ and $N$, while
the g-modes are excited at frequencies lower than both $L$ and $N$
\nocite{unn89} (Unno et al.~1989). In a chemically uniform stellar model
the pulsation modes are evenly spaced, but chemical stratification and
variations in other relevant physical quantities cause discontinuities in
$L$ and $N$ that lead to unevenly spaced modes. These deviations from
uniform spacing (in frequency or period) contain detailed information
about the interior structure of the star.

\setcounter{figure}{1}
\begin{figure}
\begin{center}
\epsfig{file=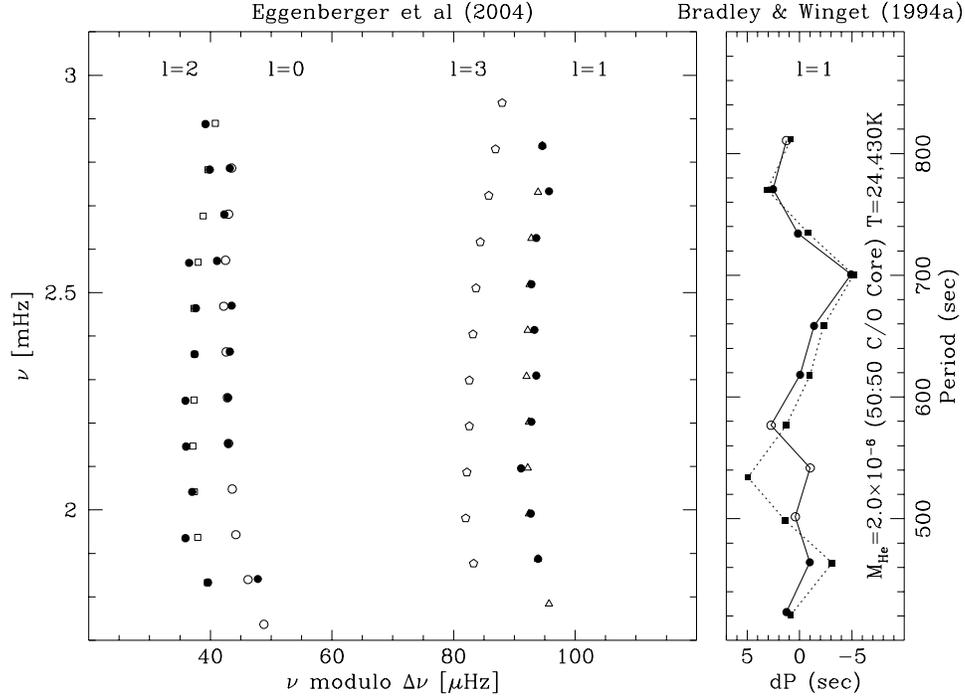,width=9.25cm,angle=270}
\caption{A recent diagram of the frequencies observed in $\alpha$~Cen~A
plotted against their deviations from the large frequency spacing (left),
and a similar diagram from ten years earlier showing the periods observed 
in GD~358 plotted against their deviations from the mean period spacing 
(right). In both cases the theoretical models match the overall patterns,
but some improvement is still needed to match the fine details.\label{fig2}}
\end{center}
\end{figure}
\nocite{bw94a,egg04}

A fundamental difference between the observed modes in Sun-like stars and
white dwarfs is that we generally see modes of only one spherical degree
in white dwarfs (mostly $\ell=1$, but in some cases $\ell=2$). In Sun-like
stars, we usually see radial ($\ell=0$) and non-radial ($\ell=1,2$ and
sometimes 3) modes excited simultaneously. Consequently, there is a
historical difference in the way the pulsation data are usually presented.
In Figure~\ref{fig2} this difference has been circumvented by rotating the
white dwarf plot. This allows us to see more readily how the current state
of model-fitting for Sun-like stars is also very similar to what was being
done with white dwarfs ten years ago.

In white dwarf models, the mean period spacing is related to the total
stellar mass, while the deviations from the mean period spacing (dP) tell
us about the thickness of the stratified surface layers \nocite{bww93}
(Bradley, Winget \& Wood 1993). In models of Sun-like stars, the large
frequency spacing ($\Delta\nu$) is related to the mean density of the
star. The deviations that split modes with odd or even spherical degree
(the so-called small frequency separation, $\delta\nu$) tell us about
chemical gradients in the interior \nocite{bg94} (see Brown \& Gilliland
1994). The current level of sophistication for model-fitting of Sun-like
stars allows us to match the overall patterns of the oscillations, but
there is room for improvement in the detailed fit to the individual modes.
This is just where white dwarf modeling was in 1994.

The story of how we improved our approach to white dwarf modeling contains
some important lessons for the modeling of Sun-like stars \nocite{met03a}
(see also Metcalfe 2003a). This is particularly relevant now, since the
next few years promise to unleash a flood of new observations of
solar-like oscillations in other stars. We have already seen what is
possible with a relatively modest space mission devoted to
asteroseismology, the Canadian MOST satellite \nocite{wal03} (Walker et
al.~2003). Forthcoming missions with larger apertures, such as COROT
\nocite{bag02} (Baglin et al.~2002) and Kepler \nocite{bor03} (Borucki et
al.~2003) are likely to extend these successes and bring new surprises.
Unfortunately, none of these missions will contribute new data on the
faintest pulsating stars, so future advances for white dwarfs and variable
sdB stars will most likely continue to come from ground-based
observations.

\section{Using Observational Constraints}

In the current model-fitting approach for Sun-like stars, it's common to
use external observational constraints on the luminosity and effective
temperature to restrict the search to a narrow range of model parameters
\nocite{dim03} (e.g., see Di~Mauro et al.~2003, Fig.~1). This guarantees
that the seismological model will be consistent with these constraints,
but it eliminates the advantage of the pulsation data. Even if we find a
reasonable fit to the pulsation modes, how do we know it couldn't be
improved if we relaxed or removed the constraints? The power of
asteroseismology is that it provides {\it independent} constraints on the
stellar structure. If the stellar models are accurate representations of
the real stars, then the pulsation modes should lead to the ``observed''
luminosity and effective temperature by themselves. If they do, this
lends additional credibility to the resulting model. If instead the
seismological data lead to a best-fit model that {\it disagrees} with the
external constraints, then something is missing from the model; And
that's exactly what we hope to learn through asteroseismology. We will
never find a disagreement if we don't look for it.

To illustrate the importance of defining the range of the search as
broadly as possible, let me tell you about the helium-atmosphere (DBV)
white dwarf GD~358. The observations of this star came from the Whole
Earth Telescope \nocite{nat90,win94} (Nather et al.~1990, Winget et
al.~1994), and the initial asteroseismological analysis was done by
\nocite{bw94a} Bradley \& Winget (1994a). For our new approach using a
parallel genetic algorithm \nocite{mc03} (Metcalfe \& Charbonneau 2003),
we decided to define the range of the search based only on the physics of
the model whenever possible.

For DBV stars, the three most important parameters are the stellar mass,
the effective temperature, and the thickness of the surface helium layer.
For our initial study we searched masses between 0.45 and 0.95~$M_\odot$.
The observed mass distribution of white dwarfs is strongly peaked near
0.6~$M_\odot$ \nocite{nap99} (Napiwotzki, Green \& Saffer 1999). Lower
mass white dwarfs are theoretically expected to have a helium core, and
take longer than the age of the universe to form through single-star
evolution. Higher mass white dwarfs are extremely rare. We searched
effective temperatures between 20,000 and 30,000~K, which easily
encompasses the full range of the DBV instability strip whether or not
trace amounts of hydrogen are allowed in the envelopes \nocite{bea99}
(Beauchamp et al.~1999). We allowed the surface helium layer to be between
$10^{-2}$ and $\sim\!10^{-8}$ in fractional mass. Thicker layers would
initiate nuclear burning at the base of the layer (ruled out by the
observations), and for thinner layers our models do not pulsate
\nocite{bw94b} (Bradley \& Winget 1994b).

\setcounter{figure}{2}
\begin{figure}
\begin{center}
\epsfig{file=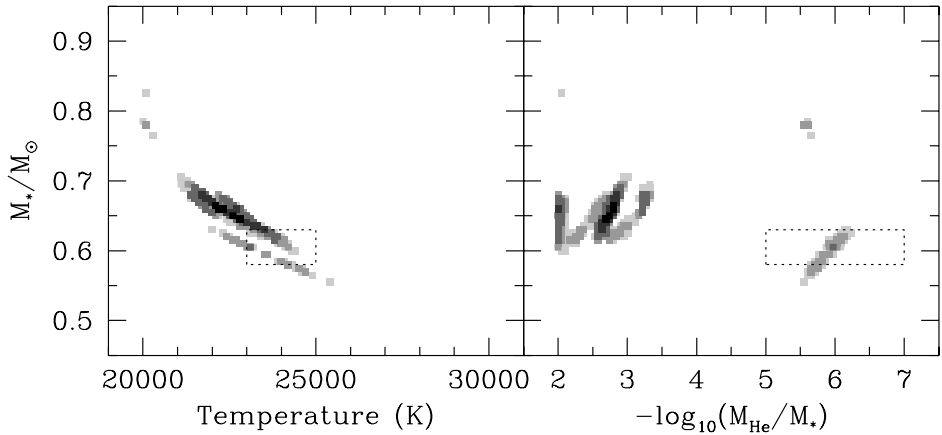,width=12.5cm,angle=0}
\caption{By searching a much broader range of parameter values compared to 
previous studies (dotted), we discovered a better match (darker) to the 
observations of GD~358. Our global search also revealed a strong sensitivity 
of the pulsations to the core composition, which eventually led to precise 
constraints on an important nuclear reaction rate.\label{fig3}}
\end{center}
\end{figure}

The full range of our search is shown in Figure~\ref{fig3}, along with the
range of the original search in 1994 (dotted). We found the same best-fit
model as Bradley \& Winget when we restricted ourselves to their search
range, but our broader search revealed an even better solution (darker)
outside of the range they considered \nocite{mnw00} (Metcalfe, Nather \&
Winget 2000). This global approach also taught us that our models are much
more sensitive to the core composition than we had previously thought.
This is significant because the oxygen mass fraction in the core of a
white dwarf is largely determined by the relative rates of the 3$\alpha$
and the $^{12}{\rm C}(\alpha,\gamma)^{12}{\rm O}$ nuclear reactions during
the red giant phase. The latter reaction is poorly constrained by
laboratory data, which are extrapolated over six orders of magnitude in
the cross-section to reach the energies relevant to red giant stars.

Motivated by our newly-discovered sensitivity to the core composition, we
added adjustable C/O profiles to our models and performed a new global
search. While our match to the observed pulsation modes improved slightly
by considering a broader range of model parameters, optimizing the C/O
profile improved the match dramatically \nocite{mwc01} (see Metcalfe,
Winget \& Charbonneau 2001, Fig.~4d). Since then we have applied the same
method to another DBV white dwarf, CBS~114, leading to two independent
constraints on the reaction rate that agree with each other \nocite{met03b} 
(Metcalfe 2003b). The two values also agree with recent extrapolations
from laboratory measurements \nocite{ang99} (Angulo et al.~1999), but are
considerably more precise.

By letting the pulsation modes speak for themselves we improved our fit to
the observations, and in the process we discovered a previously unknown
sensitivity in the models that ultimately led to constraints on an
important nuclear reaction rate. Along the way, we found inconsistencies
between some of the derived model parameters and other observations (e.g.,
spectroscopic masses and temperatures) that also prompted us to improve
the model physics. None of this would have been possible if we had defined 
our search too narrowly.

\section{Individual Modes \& Spacing}

Early attempts to detect solar-like oscillations in other stars
produced Fourier spectra that did not clearly resolve individual
frequencies, but showed excess power roughly where the excited modes were
expected. Using auto-correlation techniques, these observations could
infer the large frequency spacing, which could then be used to estimate
the mean density of the star \nocite{kje95} (see Kjeldsen et al.~1995,
Fig.~3). More recent observations can now confidently resolve the
individual modes, but auto-correlation is often still used to determine
the large frequency spacing. Because the models may suffer from systematic
errors, some recent model-fitting attempts have concentrated on the
frequency spacing rather than matching the individual frequencies. This
may be useful for a preliminary analysis, but the individual modes contain
additional information that is lost by matching only the spacing.  Using
the individual modes can also produce clues about the source of any
systematic errors, allowing us to improve the input physics of our models.

To illustrate the benefit of matching the individual modes, and not just
the mode spacing, let me tell you about our experience with the
hydrogen-atmosphere (DAV) white dwarf BPM~37093. When a white dwarf gets
cool enough, the C/O core eventually undergoes a phase transition from
liquid to solid. It crystallizes, from the center outward. This is
important because the crystallization process releases latent heat, which
delays the gradual cooling of the star. Since we can use the coolest white
dwarfs in any stellar population (the Galactic disk and halo, open and
globular clusters) to determine the age, it is crucial that we calibrate
this cooling delay. Otherwise, we will underestimate the true ages. 

A typical white dwarf with a mass of 0.6~$M_\odot$ will begin to
crystallize when it cools to an effective temperature between 6000-8000~K,
depending on the core composition. More massive white dwarfs have higher
central pressures, so they will begin to crystallize at higher
temperatures. BPM~37093 is massive enough $(\sim\!1.1~M_\odot)$ that it
has theoretically started to crystallize while at an effective temperature
inside the DAV instability strip ($\sim\!12,000$~K). The pulsations do not
penetrate the solid core, so the inner boundary for each mode is located
at the edge of the crystallized core rather than at the center of the
star. This changes the period of each mode, compared to what it would have
been in the absence of crystallization. As a consequence, we can calibrate
the theory of crystallization empirically by modeling the pulsations in
this star.

The observations of BPM~37093 were obtained by the Whole Earth Telescope
\nocite{kan00,kan05} (Kanaan et al.~2000, 2005), and the initial
theoretical study came from \nocite{mw99} Montgomery \& Winget (1999).
This early analysis concentrated on the average period spacing of the
modes, since the main effect of the crystallized core is to spread the
pulsation periods farther apart. However, they found a troubling
degeneracy between changes to the crystallized mass fraction and changes
to the thickness of the surface hydrogen layer, leading to ambiguous
results from the average period spacing (see Figure~\ref{fig4}). By
exploiting the information contained in the {\it individual periods}, we
were able to lift this degeneracy and determine a unique best-fit model.
Of course, in the small grid of models shown in Figure~\ref{fig4} we have
fixed several other important parameters, so the true best-fit model
requires a more global exploration.

\setcounter{figure}{3}
\begin{figure}
\begin{center}
\epsfig{file=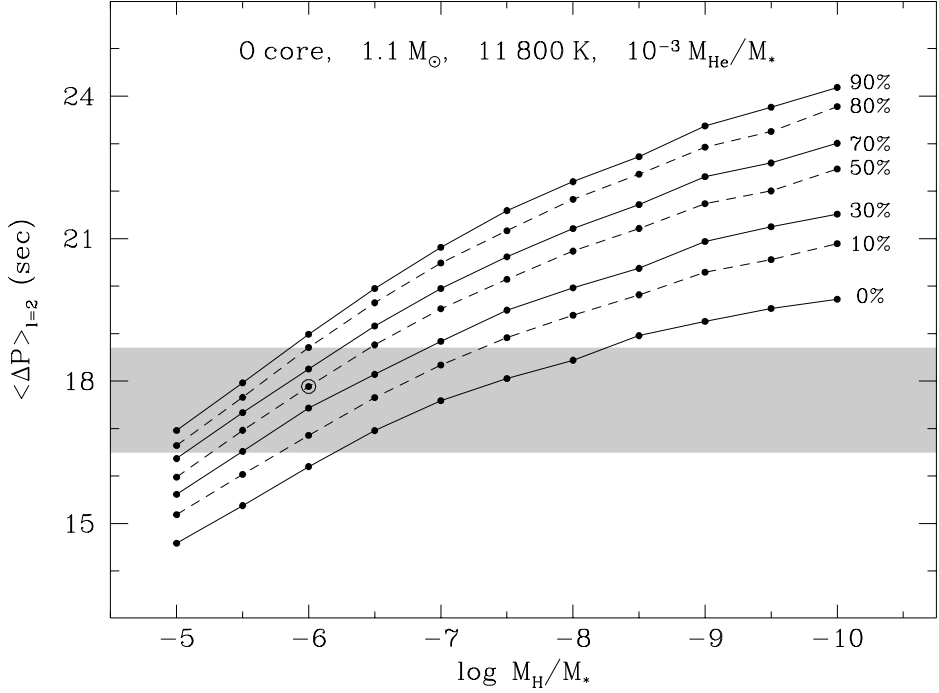,width=9.25cm,angle=270}
\caption{When our fit used only the {\it mode spacing} (shaded), we were 
unable to distinguish between thinner H layers and a larger crystallized 
mass fraction in BPM~37093---both parameters increased the mode spacing. 
By exploiting the information in the {\it individual modes}, one model 
(circled) stands out as the best in this small grid. Of course, a 
simultaneous optimization of all of the parameters is needed to find the 
global solution.\label{fig4}}
\end{center}
\end{figure}

This is a huge computational problem, since we need to consider the
stellar mass, the effective temperature, the thickness of the helium
mantle and the surface hydrogen layer, the core composition, and the
crystallized mass fraction, all simultaneously. Our initial fits to the
individual periods confirmed that BPM~37093 is substantially crystallized
\nocite{mmk04} (Metcalfe, Montgomery \& Kanaan 2004), but the search was
too limited to determine the exact fraction. Our latest results look
promising \nocite{mmk05} (Metcalfe, Montgomery \& Kanaan 2005), and have
recently been confirmed by \nocite{bf05} Brassard \& Fontaine (2005). But
this expanded search still required that we fix the mass and core
composition, and there remain many locally optimal models, so our search
for a global solution continues.

By using the individual modes, and not just the spacing, we lifted the
degeneracy between the crystallized mass fraction and the surface hydrogen
layer mass in models of the DAV white dwarf BPM~37093. This led to the
first direct confirmation of crystallization theory, and opened the door
to a more global exploration of the models which is still in progress. If
we had confined ourselves to fitting the average period spacing, we would
never have obtained this first glimpse inside of a crystallized star.

\section{Summary}

Let me summarize the main points I hope you will take away with you:
\begin{itemize}
\item The future of asteroseismology for Sun-like stars is primarily in 
      space.
\end{itemize}
We soon expect many new observations from current and planned satellite 
missions. It will be a distinct advantage to have in place the analysis 
methods that will help us make sense of these data efficiently.
\begin{itemize}
\item \underline{Lesson 1}: Let the pulsations speak for themselves.
\end{itemize}
To interpret these data, it is important that we avoid introducing our
subjective biases by defining the range of our model-fitting search too 
narrowly. The asteroseismic models should agree with independent 
measurements without coercion.
\begin{itemize}
\item \underline{Lesson 2}: Use the individual modes, not just the 
      spacing.
\end{itemize}
The individual frequencies contain additional information that is lost if 
we focus only on the overall pattern of frequency spacing. This could 
unnecessarily limit what we are ultimately able to derive from the 
observations.
\begin{itemize}
\item The future of asteroseismology for fainter stars is multi-site.
\end{itemize}
Upcoming space missions will not affect the future of the fainter 
pulsators. Advances in our understanding of white dwarfs and variable 
subdwarf B stars will continue to come from multi-site campaigns, possibly 
with larger telescopes.

This is a very exciting time for asteroseismology. It is the beginning of
a new era, when we will improve our knowledge of solar physics by studying
Sun-like stars. This will allow us to explore the past and future of our
own Sun by observing pulsations in solar analogues, calibrating our models
at different evolutionary stages and under various physical conditions.
For the white dwarf stars, we are closer to the end of an era. Over the
past decade we have steadily improved our models and our analysis methods,
charting a course that is just beginning for Sun-like stars. If we are
clever enough, we can use what we have learned along the way to accelerate
our progress on other types of stars.\\

\noindent
{\it Acknowledgments.} This work was supported by the National Science 
Foundation through an Astronomy \& Astrophysics Postdoctoral Fellowship.
I would like to thank J{\o}rgen Christensen-Dalsgaard for motivating me
to turn my attention to Sun-like stars, and Tim Brown for helping to bring 
this project to the High Altitude Observatory at the National Center for
Atmospheric Research.

\end{document}